\documentclass{elsart1p}
\usepackage{graphicx}
\usepackage{amssymb}
\begin{document}
\begin{frontmatter}
%
%
%
\title{Recent Results and Future Prospects From MINOS}
%
%
\author{Jonathan M. Paley, for the MINOS Collaboration}
\address{Indiana University Physics Department, Bloomington, IN  47403, USA}
\begin{abstract}
The MINOS experiment uses the intense NuMI beam created at Fermilab and two
magnetized tracking calorimeters, one located at Fermilab and one located 
735 km away at the Soudan Mine in Minnesota, to make precise measurements of 
$\nu_\mu$ disappearance oscillation parameters.  We present recent results 
from the first two years of NuMI beam operations, including the precise 
measurement of the atmospheric neutrino oscillation parameters and the search 
for sterile neutrinos.  Future prospects for MINOS will also be discussed, 
including an improved limit on the $\theta_{13}$ mixing angle by searching 
for $\nu_e$ appearance in the $\nu_\mu$ beam.
\end{abstract}
\begin{keyword}
%
\PACS 14.60.Pq \sep 14.60.St
\end{keyword}
\end{frontmatter}
%
The Main Injector Neutrino Oscillation Search (MINOS) experiment was designed 
to primarily confirm neutrino oscillations by allowing a precise measurement 
of the atmospheric neutrino oscillation parameters $\Delta m^2$ and 
$\sin^2(2\theta)$.  MINOS is also capable of searching for 
sterile neutrinos and the subdominant $\nu_\mu \rightarrow \nu_e$ oscillation.

MINOS utilizes the intense Neutrinos at the Main Injector (NuMI) beam at Fermi 
National Laboratory.  A very pure beam of muon neutrinos is aimed at an 
underground laboratory 735 km from the NuMI production target in Soudan, MN, 
where a 5.4 kton magnetizied iron tracking calorimeter (the Far Detector, FD) 
is used to detect the neutrinos\cite{MINOS:NIM}.  A functionally identical 0.98 
kton Near Detector (ND) is located approximately 1 km downstream from the NuMI 
target.

\begin{figure}[ht]
\begin{minipage}[t]{0.48\linewidth}
\centering
\includegraphics*[width=70mm,height=60mm]{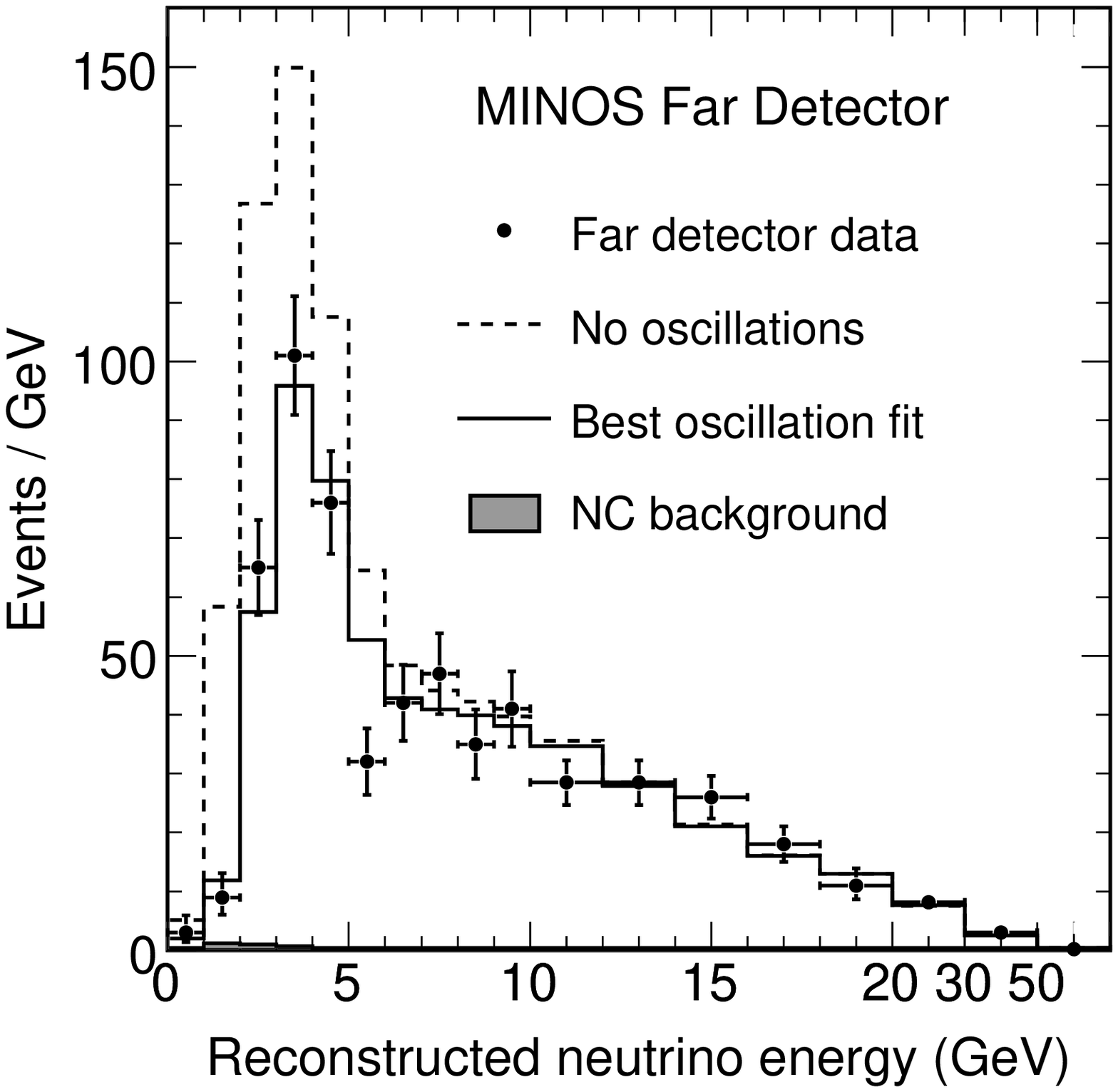}
\caption{Measured $\nu_\mu$ energy spectrum in the MINOS FD (black points), compared to the expected spectrum for the case of no oscillations.}
\label{fig:Spectrum}
\end{minipage}
\hspace{0.02\linewidth}
\begin{minipage}[t]{0.48\linewidth}
\centering
\includegraphics*[width=70mm,height=60mm]{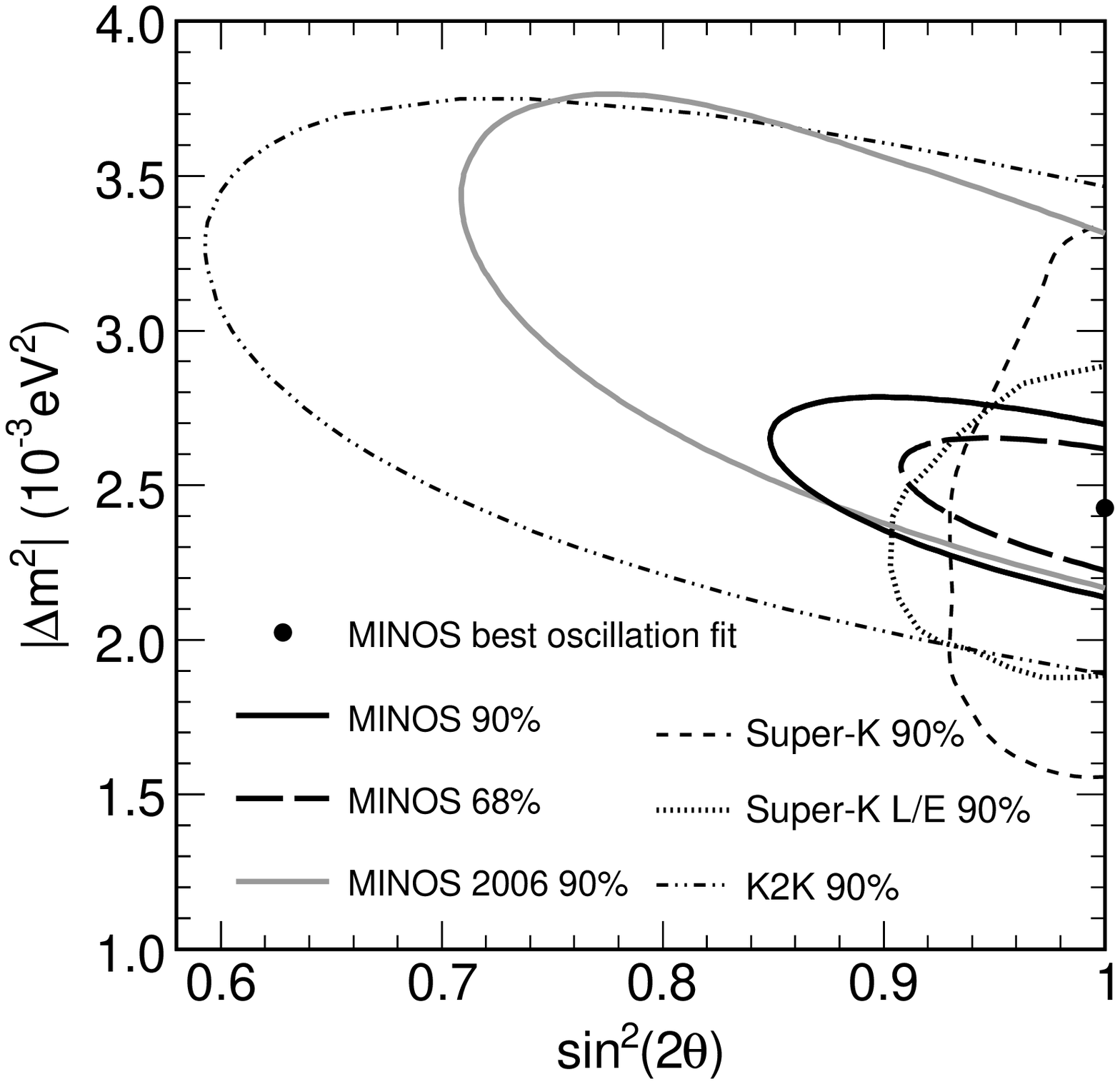}
\caption{68\% and 90\% contour lines for the oscillation fit parameters $\Delta m^2$ and $\sin^2(2\theta)$ compared to other experimental measurements.}
\label{fig:Contour}
\end{minipage}
\end{figure}

The measurement of the atmospheric neutrino oscillation parameters 
$\Delta m^2$ and $\sin^2(2\theta)$ is accomplished by 
comparing the measured muon neutrino energy spectra in the ND and FD.
Charged-current (CC) interaction events are separated from neutral-current 
(NC) events in the detectors based on topological 
characteristics that indicate a muon in the final state: track length, mean 
pulse height, fluctuation in pulse height and transverse track profile.  A CC/NC
separation parameter cut is determined that maximizes the CC event selection 
efficiency and minimizes the NC background\cite{MINOS:CCNumu}.  

The energy spectrum for CC events is measured in the ND and extrapolated to the 
FD.  The extrapolation method uses MC to determine energy smearing and 
acceptance corrections for the expected FD energy spectrum.  The dominant 
systematic uncertainties in the predicted FD spectrum are a 4\% 
normalization uncertainty, a 10.3\% hadronic energy calibration uncertainty 
and a 50\% uncertainty on the NC background that pass the CC cut selection 
criteria.  The normalization and hadronic energy uncertainties primarily effect
the measurement of $\Delta m^2$ at the level of 
$\pm 0.081 \times 10^{-3}$ eV$^2$ and $\pm 0.052 \times 10^{-3}$ eV$^2$ 
respectively.  The NC background is the largest effect on the measurement of 
$\sin^2(2\theta)$ at the level of $\pm0.016$.

Fig.~\ref{fig:Spectrum} shows 
the measured FD energy spectrum, the expected spectrum with no oscillations, 
the best oscillation fit to the data, and the NC background (significant only 
in the lowest energy bin).  A total of 848 events are observed in the FD, with 
$1065 \pm 60$ events expected under the no-oscillation hypothesis for a data 
set based on $3.36\times 10^{20}$ protons on target (POT).  Fitting the
observed energy spectrum to the survival probability, and constraining the
fit parameters to their physically meaningful values, we find 
$\Delta m^2 = 2.43 \pm 0.13 \times 10^{-3}$ and 
$\sin^2(2\theta) > 0.90$ (90\% CL), with a $\chi^2/ndof = 90/97$.  
Fig.~\ref{fig:Contour} shows the oscillation parameter phase space allowed by 
the latest MINOS measurement, with comparisons to other experiments and the 
previous MINOS measurement.  Alternative hypotheses of neutrino 
decay\cite{Decay:Barget} and neutrino decoherence\cite{Deco:Fogli} have also 
been tested with the MINOS data, and these are disfavored with respect to the 
osciallation hypothesis by $3.7\sigma$ and $5.7\sigma$ respectively.

\begin{figure}[t]
\begin{minipage}[t]{0.48\linewidth}
\centering
\includegraphics*[width=75mm]{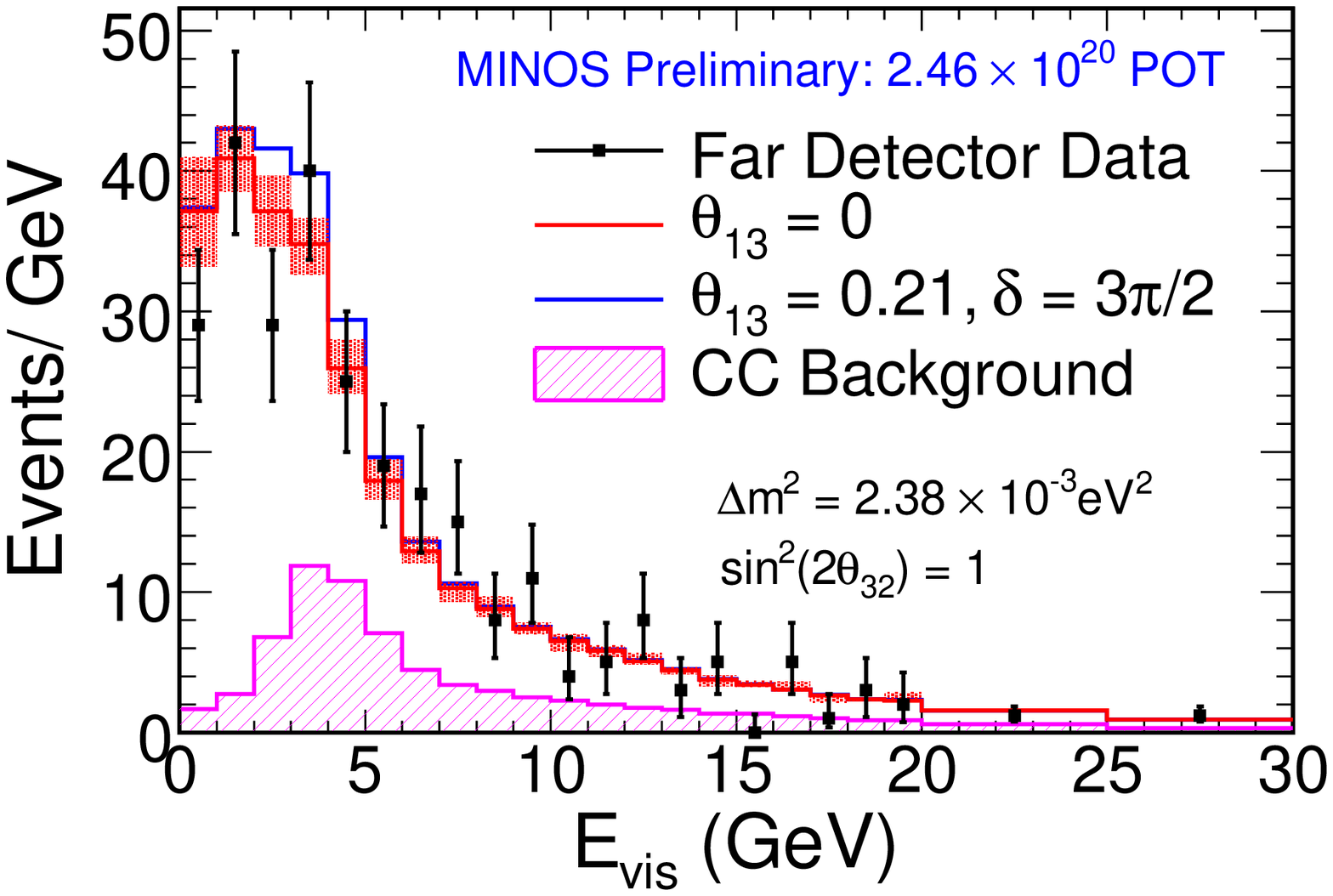}
\caption{Visible energy spectrum for NC events in the FD compared to the 
dominant CC $\nu_\mu$ background and predicted spectra for $\theta_{13}$=0 
and $\theta_{13}$ at the Chooz limit.}
\label{fig:NCSpect}
\end{minipage}
\hspace{0.02\linewidth}
\begin{minipage}[t]{0.48\linewidth}
\centering
\includegraphics*[width=65mm]{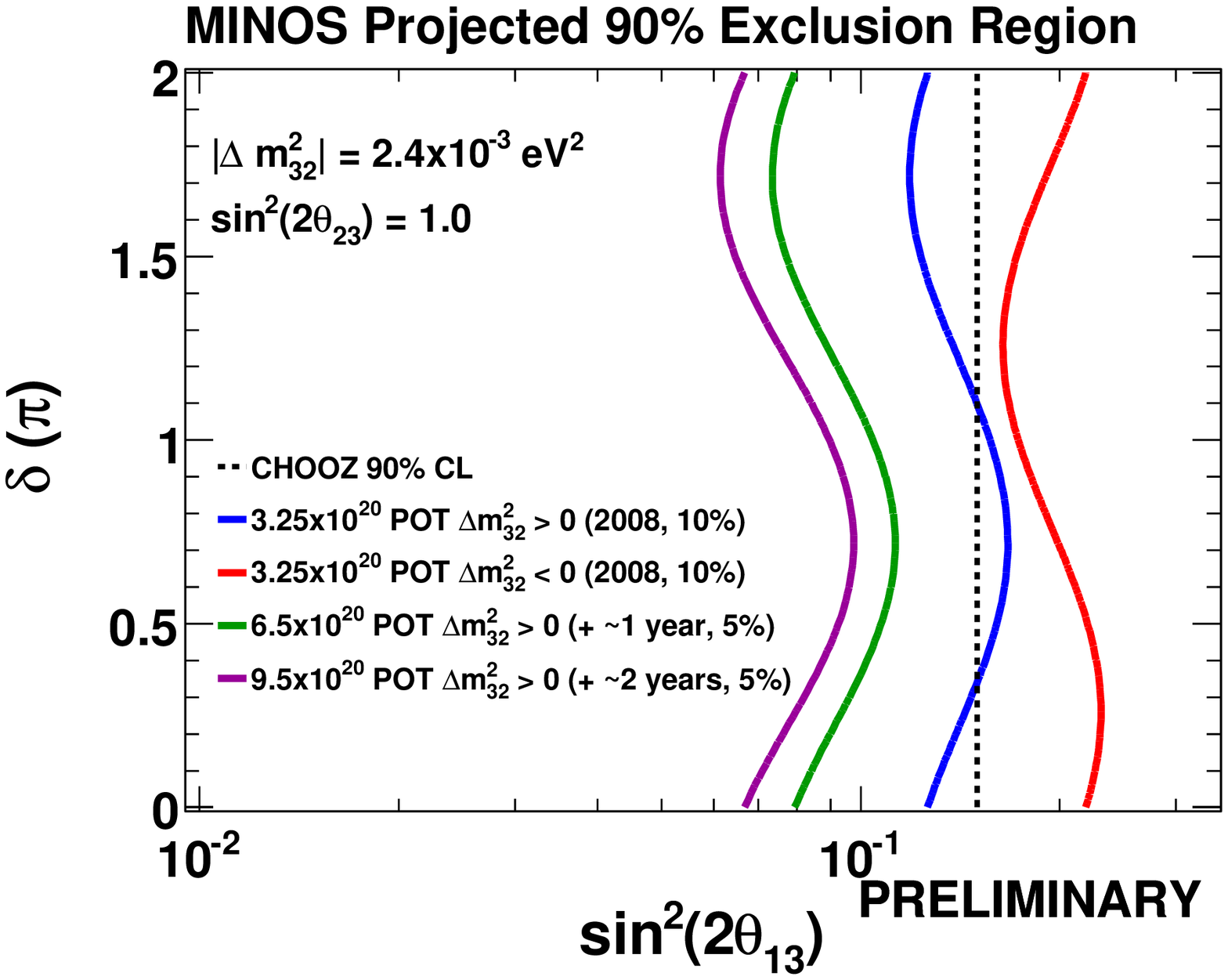}
\caption{MINOS sensitivity to $\sin^2(2\theta_{13})$ for various data sets. 
The current data set will produce a sensitivity roughly equal to the current 
best limit (solid black line).}
\label{fig:Nue}
\end{minipage}
\end{figure}

Since the NC event rate is independent of neutrino flavor, and thus unaffected 
by oscillations between the three active neutrino flavor states, a deficit in
the NC rate in the FD would indicate the existence of at least one additional 
sterile neutrino ($\nu_s$).  The visible energy spectrum of NC (shower-like) 
events in the ND is extrapolated to the FD in search for such a deficit.

Fig.~\ref{fig:NCSpect} shows the measured visible energy spectrum in the FD,
based on $2.46 \times 10^{20}$ POT,
along with the expected spectra for $\theta_{13}$=0 and $\theta_{13}$ at the 
Chooz limit\cite{Chooz} and the predicted $\nu_\mu$ CC event background.
The quantity 
$f_s = (P_{\nu_\mu \rightarrow \nu_s})/(1 - P_{\nu_\mu \rightarrow \nu_\mu})$ describes
the fraction of $\nu_\mu$ that have oscillated to $\nu_s$ in a simple 
four-flavor model where oscillations to sterile neutrinos ocurr at the same 
mass splitting as the $\nu_\mu$ disappearance measured from the CC interactions.
For the case of $\theta_{13}$=0, $f_s = 0.28^{+0.25}_{-0.28}$ (stat. + syst.) 
or $f_s < 0.68$ (90\% CL).  For the case of $\theta_{13}$ at the Chooz limit, 
$f_s = 0.43^{+0.23}_{-0.27}$ (stat. + syst.) or $f_s < 0.80$ (90\% CL) 
\cite{MINOS:NC}.

An excess of $\nu_e$ events in the MINOS FD would indicate 
$\nu_\mu \rightarrow \nu_e$ oscillations and a non-zero value for the 
$\theta_{13}$ oscillation term.  CC $\nu_e$ events in the MINOS detectors
are tagged by searching for events that have electromagnetic shower profiles
in the final state.  The CC $\nu_e$ events appear very similar to NC events 
which are the dominant background, although low-energy CC $\nu_\mu$ events 
also contribute significantly to the background in this measurement.
For a data set based on $3.25\times 10^{20}$ POT, approximately 12 CC 
$\nu_e$ events are expected in the FD with 42 background events.  


Fig.~\ref{fig:Nue} shows the MINOS sensitivity to $\sin^2(2\theta_{13})$ as a 
function of the CP violating phase $\delta$.  With the current data set being 
analyzed based on $3.25\times 10^{20}$ POT, the sensitivity of the MINOS 
measurement is roughly that of the current best limit; within about a year, we 
expect an improvement in the limit by about a factor of two based on a 
doubling of statistics and a reduction by 50\% of the estimated backgrounds.
We expect to present first results on the search for 
$\nu_e$ appearance in the MINOS detectors by spring of 2009.

%
%
%

%
\end{document}